\documentclass[showpacs,aps]{revtex4}
\usepackage{graphicx}

\newcommand{\ep}{\epsilon}

\newcommand{\pa}{\partial}
\newcommand{\td}{\tilde}

\begin{document}

\title{Isospin Symmetry Breaking in $\rho\rightarrow\pi\gamma$ Decay
\footnote{This work is partially supported by NSF of China through
C.N. Yang.}}
\author{Xiao-Jun Wang\footnote{wangxj@mail.ustc.edu.cn}}
\affiliation{Institution of theoretical physics, BeiJing, 100080,
P.R. China}
\author{Ji-Hao Jiang}
\affiliation{Center for Fundamental Physics,
University of Science and Technology of China\\
Hefei, Anhui 230026, P.R. China}
\author{Mu-Lin Yan}
\affiliation{  CCAST(World Lab), P.O. Box 8730, Beijing, 100080, P.R. China \\
  and\\
 Center for Fundamental Physics,
University of Science and Technology of China\\
Hefei, Anhui 230026, P.R. China\footnote{mail
address}}

\begin{abstract}
In terms of effective field theory and mixed propagator approach,
we show that there is a larger hidden effect of isospin breaking
in $\rho^0\rightarrow\pi^0\gamma$ decay due to an $\omega$
exchange, $\rho^0\rightarrow\omega\rightarrow\pi^0\gamma$. The
branching ratio is predicted as
$B(\rho^0\rightarrow\pi^0\gamma)=(11.67\pm 2.0)\times 10^{-4}$,
which is much larger than Partcile Data Group's datum $(6.8\pm
1.7)\times 10^{-4}$ and one of charged mode,
$B(\rho^\pm\rightarrow\pi^\pm\gamma)=(4.5\pm 0.5)\times 10^{-4}$.
\end{abstract}
\maketitle

\section{Introduction}

In hadron physics, the isospin symmetry or charge symmetry\cite{Miller90}
is broken by inequality of the light quark masses, especially $m_u\neq
m_d$, and electromagnetic interaction of hadrons. This breaking of isospin
symmetry induces various measurable physics effects such as $\pi^0-\eta$,
$\Lambda-\Sigma^0$ mixing, $\omega\rightarrow\pi^+\pi^-$ decay, etc. It is
common knowledge that the effects of isospin breaking can be omittd in the
isospin conservation processes. However, this argument is not always
right. The purpose of this letter is to show that a large isospin breaking
effect is in isospin conservation decay $\rho^0\rightarrow\pi^0\gamma$
(so-called hidden isospin symmetry breaking effect). Thus the real
branching ratio for this decay should be much larger than datum cited by
PDG-2000\cite{PDG2000} and charged mode.

The anomalous-like radiative decays of light flavour vector
mesons have been observed by several group: The branching ratio
for charged $\rho$ is $B(\rho^\pm\rightarrow\pi^\pm\gamma)=(4.5\pm
0.5)\times 10^{-4}$\cite{ccho}. The branching ratios for $\rho^0$
and $\omega$ decays were first obtained in
refs.\cite{Dru84,Dolinsky89,Dolinsky91} by using the data of
$e^+e^-$ collider in Neutral Detector, but the triangle anomaly
contribution of QCD was ignored in their analysis. Benayoun {\sl
et al.} has reanalyzed the $\rho^0$ and $\omega$ decay to a
pseudoscalar meson plus a photon via taking into account the
triangle anomaly contribution\cite{Benayoun96}. The authors use
two models ($M_1$ and $M_2$)\cite{Benayoun93,Benayoun95} to fit
data. It is a bit surprising that they found two local minima for
both two model fits, one with $\chi^2/N_{df}\simeq 0.5$ (solution
A) and another one with $\chi^2/N_{df}\simeq 0.7$ (solution B).
To see table 1, where phase $\Phi_V$ of vector mesons is defined
via $e^+e^-\rightarrow\pi^0\gamma$ cross section (equation (1) of
ref.\cite{Benayoun96}). Although the solution A has smaller
$\chi^2/N_{df}$, and the phase difference $\Phi_\phi-\Phi_\rho$
is found around $210^{\circ}$ in the solution A, close to
expectation from the quark model ($180^{\circ}$), the $\rho^0$
branching ratio is much larger than expected from the charged mode
or from SU(3) symmetry. Thus the authors took the solution B as
final result for agreeing with isospin symmetry. However, this
conclusion is incorrect. In this paper let us see what happens
there.

The paper is organized as follows: In section 2, we first give the
formula of the transition amplitude for
$\rho,\;\omega\to\pi\gamma$. Then we will prove this formula via
two independent methods: effective field theory approach and mixed
propagator approach. In section 3, we will provide final numeric
results, and a brief conclusion.

\begin{table}[tbp]
\centering
\caption{Branching ratios and phases obtained from fit to
$e^+e^-\rightarrow\pi^0\gamma$ cross section using two $\rho^0$ models and
assuming $\Phi_\rho=\Phi_\omega$. The table is abstracted from table 4 of
ref.[7].}
 \begin{tabular}{cccccc}\hline
$\Gamma(\pi^0\rightarrow\gamma\gamma)$=7.7eV
&\multicolumn{2}{c}{Model $M_1$}&
&\multicolumn{2}{c}{Model $M_2$} \\ \cline{2-3}\cline{5-6}
& sol. A& sol. B &&sol. A&sol. B\\ \hline
$\rho^0\rightarrow\pi^0\gamma$ (units $10^{-4}$)& $(11.51\pm 2.00)$&
$(6.17\pm 1.57)$&&$(11.67\pm 2.00)$& $(6.77\pm 1.72)$\\
$\omega\rightarrow\pi^0\gamma$ (units $10^{-2}$)& $(8.62\pm 0.26)$&
$(8.39\pm 0.24)$&&$(8.64\pm 0.27)$& $(8.39\pm 0.24)$\\
$\phi\rightarrow\pi^0\gamma$ (units $10^{-3}$)& $(1.15\pm 0.13)$&
$(1.20\pm 0.16)$&&$(1.21\pm 0.16)$& $(1.26\pm 0.17)$\\
$\Phi_{\rho}=\Phi_{\omega}$ [degrees]&$-94\pm 8$ &$124\pm 9$&
&$-90\pm 7$&$125\pm 11$\\
$\Phi_\phi$ [degrees]&$114\pm 14$ &$248\pm 9$&
&$119 {+11\atop -18}$&$248{+18\atop -10}$ \\
$\chi^2$/dof&29/57&38/57&&29/57&36/57 \\ \hline
   \end{tabular}
\begin{minipage}{5in}
\end{minipage}
\end{table}

\section{Transition amplitude for $V\to \pi\gamma$}

The transition matrix element for vector mesons decay to a pion and a
photon is
\begin{eqnarray}\label{1}
<\pi(k)\gamma(q_1)|V(q_2)>=if_{V\pi\gamma}\ep^{\mu\nu\alpha\beta}
q_{1\mu}\ep_\nu^*q_{2\alpha}e_\beta,
\end{eqnarray}
where $e_\mu$ and $\ep_\mu$ are polarized vector of vector mesons and
photon respectively. At the large $N_c$ limit and chiral limit, if we
ignore possible contribution from resonance exchange, the exact isospin
symmetry implies
\begin{eqnarray}\label{2}
f_{\rho^\pm\pi^\pm\gamma}=f^{(0)}_{\rho^0\pi^0\gamma}
=\frac{1}{3}f^{(0)}_{\omega\pi^0\gamma},
\end{eqnarray}
where the superscript $``(0)''$ denotes absence of the resonance exchange.
A complete consideration for $\rho^0$ and $\omega^0$ decays should
include the contribution from the following resonance exchange,
\begin{eqnarray}\label{3}
f_{\rho^0\pi^0\gamma}^{(c)}&=&f^{(0)}_{\rho^0\pi^0\gamma}
  +\frac{\Pi_{\rho\omega}(m_\rho^2)f^{(0)}_{\omega\pi^0\gamma}}
   {m_\rho^2-m_\omega^2+im_\omega\Gamma_\omega},\nonumber\\
f_{\omega\pi^0\gamma}^{(c)}&=&f^{(0)}_{\omega\pi^0\gamma}
  +\frac{\Pi_{\rho\omega}(m_\omega^2)f^{(0)}_{\rho^0\pi^0\gamma}}
   {m_\omega^2-m_\rho^2+im_\rho\Gamma_\rho},
\end{eqnarray}
where the momentum-dependent $\rho^0-\omega$ mixing amplitude
$\Pi_{\rho\omega}(q^2)$ is defined in the following effective
lagrangian. The relations in equation~(\ref{3}) holds more generally. In
the other words, the couplings $f^{(0)}_{\rho^0\pi^0\gamma}$ and
$f^{(0)}_{\omega\pi^0\gamma}$ can include corrections beyond large $N_c$
limit and chiral limit. Meanwhile, the relation~(\ref{2}) will be broken,
but this breaking is slight. This point can be checked by using datum of
$\rho^\pm\to\pi^\pm\gamma$ decay and $\omega\to\pi^0\gamma$ decay if we
believe $f_{\rho^\pm\pi^\pm\gamma}=f^{(0)}_{\rho^0\pi^0\gamma}$.

It must be point out that the equation~(\ref{3}) is a
non-perurbative results instead of a perturbative expression. In
the rest of this section we will prove it by two independent
methods: effective field theory approach and mixed propagator
approach, respectively.

\subsection{Effective field theory approach}

The most general effective lagrangian concerning
$\rho^0\rightarrow\pi^0\gamma$ and $\omega\rightarrow\pi^0\gamma$
decays as follow
\begin{eqnarray}\label{a1}
{\cal L}={\cal L}_0+{\cal L}_I+{\cal L}_2+{\cal L}_{\rm ChPT}
+{\cal L}_{\rm WZW},
\end{eqnarray}
where ${\cal L}_{\rm ChPT}$ is lagrangian of chiral perturbative
theory, ${\cal L}_{\rm WZW}$ is Wess-Zumino-Witten lagrangian,
${\cal L}_0$ and ${\cal L}_I$ are lagrangians of free fields and
interaction for vector mesons respectively, ${\cal L}_2$ is
counterterm lagrangian for ${\cal L}_0$ and ${\cal L}_I$
\footnote{An effective theory is renormalized if number of
coupling constant in the effective lagrangian is infinite, e.g.,
Chiral Perturbative Theory. Here we focus on the most general
effective lagrangian, thus it can be treated as renormalized
one.}. Explicitly, ${\cal L}_0$ and ${\cal L}_I$ can be written as
follows
\begin{eqnarray}\label{a2}
{\cal L}_0&=&-\frac{1}{4}(\pa_\mu\rho_\nu^i-\pa_\nu\rho_\mu^i)
    (\pa^\mu\rho^{i\nu}-\pa^\nu\rho^{i\mu})
-\frac{1}{4}(\pa_\mu\omega_\nu-\pa_\nu\omega_\mu)
    (\pa^\mu\omega^{\nu}-\pa^\nu\omega^{\mu})\nonumber \\&&\hspace{0.3in}
 +\frac{1}{2}\td{m}_\rho^2\rho_\mu^i\rho^{i\mu}+\frac{1}{2}
 \td{m}_\omega^2\omega_\mu\omega^\mu+\cdots, \nonumber \\
{\cal L}_I&=&\int\frac{d^4q}{(2\pi)^4}e^{iq\cdot x}{\Big \{}
 \Pi_{\rho\omega}(q^2)(g_{\mu\nu}-\frac{q_\mu q_\nu}{q^2})
 \omega^\mu(x)\rho^{0\nu}(q)+i\ep^{ijk}f_{\rho\pi\pi}(q^2)
 \rho_i^\mu(q)\pi_j(x)\pa_\mu\pi_k(x)\nonumber\\ &&\hspace{0.3in}
 +\ep^{\mu\nu\alpha\beta}\ep^{ijk}f_{\omega3\pi}(q^2)
  \omega_\mu(q)\pa_\nu\pi_i(x)\pa_\alpha\pi_j(x)\pa_\beta\pi_k(x)
 +\ep^{\mu\nu\alpha\beta}f_{\rho\pi\gamma}(q^2)
  q_\mu\rho_\nu^i(q)\pa_\alpha A_\beta(x)\pi_i(x)
   \nonumber\\ &&\hspace{0.3in}
 +\ep^{\mu\nu\alpha\beta}f_{\omega\pi\gamma}(q^2)
  q_\mu\omega_\nu(q)
  \pa_\alpha A_\beta(x)\pi^0(x)+\cdots{\Big \}}.
\end{eqnarray}
Several remarks are related to lagrangian~(\ref{a2}):
\begin{enumerate}
\item Focusing on any low energy effective theory, a general knowledge is
that any coupling should be momentum-dependent\cite{WY00}. In
lagrangian~(\ref{a2}) this truth has been exhibited via
momentum-dependence of form factors $f_{\rho\pi\pi}(q^2)$,
$f_{\omega3\pi}(q^2)$, etc.. All these form factors should be real
function of vector meson
four-momentum square, $q^2$. So that the unitarity of $S$-matrix prevents
us from taking complex mass square in on-shell transition amplitude.
\item $\td{m}_\rho$ and $\td{m}_\omega$ in ${\cal L}_0$ are not physical
masses of $\rho^0$ and $\omega$, since they will be shifted due to
$\rho^0-\omega$ mixing~
\footnote{Of course, there should be other possible mechanism which can
also cause mass splitting of $\rho^0$ and $\omega$. In this paper,
however, we only provide a heuristical discussion. Thus we ignore other
ingredients here.}.
The physical masses of $\rho^0$ and $\omega$ correspond to poles in their
type I complete propagators (fig. 1)
\footnote{Due to effective lagrangian~(\ref{a2}), the complete propagators
of $\rho^0$ and $\omega$ receive two kinds of contribution. One is from
$\rho^0-\omega$ mixing and another is from meson loops. We label them as
type I and type II contribution respectively. In addition, the
renormalization does not shift the pole of propagator, thus the pole in
type I complete propagators is same to one in complete propagators
considering both two kinds of contributions.},
in which we have
considered contribution from $\rho^0-\omega$ mixing completely:
\begin{eqnarray}\label{a3}
\Delta_{(\rho)\mu\nu}^{\rm (I)}(q^2)=\frac{-ig_{\mu\nu}}
  {q^2-\td{m}_\rho^2-\Pi_{\rho\omega}(q^2)},\nonumber \\
\Delta_{(\omega)\mu\nu}^{\rm (I)}(q^2)=\frac{-ig_{\mu\nu}}
  {q^2-\td{m}_\omega^2+\Pi_{\rho\omega}(q^2)}.
\end{eqnarray}

\begin{figure}[hptb]
\centering
   \includegraphics[width=3.5in]{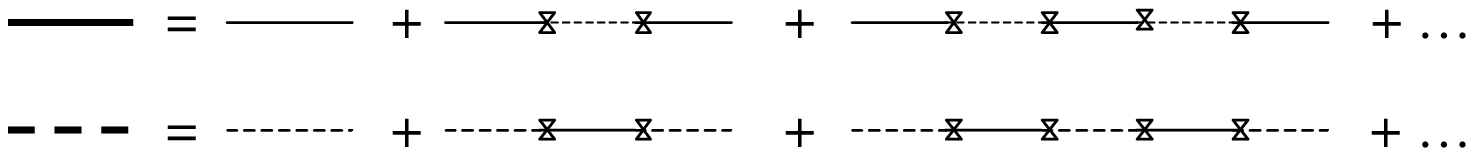}
\begin{minipage}{5in}
\caption{Non-perturbative correction of $\rho^0-\omega$ mixing to
propagators of $\rho^0$ and $\omega$. The physical masses of $\rho^0$ and
$\omega$ correspond to the poles of type I complete propagators. Here
solid
lines and dash lines denote nonphysical propagator of $\rho^0$ and
$\omega$ respectively, and bold solid line and bold dash line denote type
I complete propagators of $\rho^0$ and $\omega$ respectively.}
\end{minipage}
\end{figure}

These type I complete propagators give the physical masses of $\rho^0$ and
$\omega$ as follows
\begin{eqnarray}\label{a4}
m_\rho^2=\td{m}_\rho^2+\Pi_{\rho\omega}(m_\rho^2),\hspace{0.5in}
m_\omega^2=\td{m}_\omega^2-\Pi_{\rho\omega}(m_\omega^2).
\end{eqnarray}
It is well-known that physical mass obtained from the pole of complete
propagator should be equal to one obtained from orthogonal rotating
$\rho^0$ and $\omega$ to their mass eigenstates.

\item When we write interaction lagrangian~(\ref{a2}), we have
conveniently assumed that widths of $\rho$ and $\omega$ are generated
dynamically by pion loops. This assumption can be released, and it
does not affect the following formal discussion.
\end{enumerate}

In Heisenberg picture, the transition matrix~(\ref{1}) can be expressed in
terms of LSZ reducation formula\cite{LSZ55}
\begin{eqnarray}\label{a5}
&&<\pi(k)\gamma(q_1),out|V(q_2),in> \nonumber \\
&=&i^3(Z_3^{(\rho)}Z_3^{(\pi)}
 Z_3^{(\gamma)})^{-{1\over 2}}\int d^4xd^4yd^4z
 \frac{e^{-ik\cdot x}}{\sqrt{(2\pi)^32\omega_{\vec{\bf k}}}}
 \sum_{\lambda,\sigma}
 \frac{\ep^\mu_{\vec{\bf q}_1\lambda}e^{-iq_1\cdot y}}
 {\sqrt{(2\pi)^32\omega_{\vec{\bf q}_1}}}
 \frac{e^\nu_{\vec{\bf q}_2\sigma}e^{-iq_2\cdot z}}
 {\sqrt{(2\pi)^32\omega_{\vec{\bf q}_2}}}\nonumber \\ && \times
 (\pa_x^2+m_\pi^2)\pa_y^2(\pa_z^2+m_\rho^2)
 (0|T\{\pi(x)A_\mu(y)V_\nu(z)\}|0)_{\rm H},
\end{eqnarray}
where $Z_3$ are renormalization constant of wave function, and the
subscript ``H'' denotes in Heisenberg picture. The above expression can be
transformed to interaction picture via
\begin{eqnarray}\label{a6}
(0|T\{\pi(x)A_\mu(y)V_\nu(z)\}|0)_{\rm H}
=<0|T\{\pi(x)A_\mu(y)V_\nu(z)e^{i\int d^4x'({\cal L}_I
+{\cal L}_{\rm ChPT}+{\cal L}_{\rm WZW})(x')}\}|0>_{\rm I},
\end{eqnarray}
where the subscript ``I'' denotes in interaction picture.

In the language of Feymann diagram, the non-perturbative results can be
obtained by summing all diagrams of perturbative expansion (fig. 2).
Here we focus on $\rho^0\rightarrow\pi^0\gamma$ decay, and the discussion
for $\omega\rightarrow\pi^0\gamma$ is similar. In fig.~2, every
``$\bullet$'' denotes contribution from all potential meson loops. Thus
the renormalization of mass and wave function of $\rho^0$ and $\omega$ is
necessary. It should be pointed out that every bold solid line or bold
dash line
\footnote{We should distinguish bold bold solid line from
bold solid line in fig.~2. The former is defined as complete propagator
of $\rho^0$ while the latter is defined as type I complete propagator of
$\rho^0$},
in fig. 2 is defined in fig. 1. It means that the effect of
$\rho^0-\omega$ mixing has been included completely via summing all
diagrams of fig. 2.

\begin{figure}[hptb]
\centering
   \includegraphics[width=3.5in]{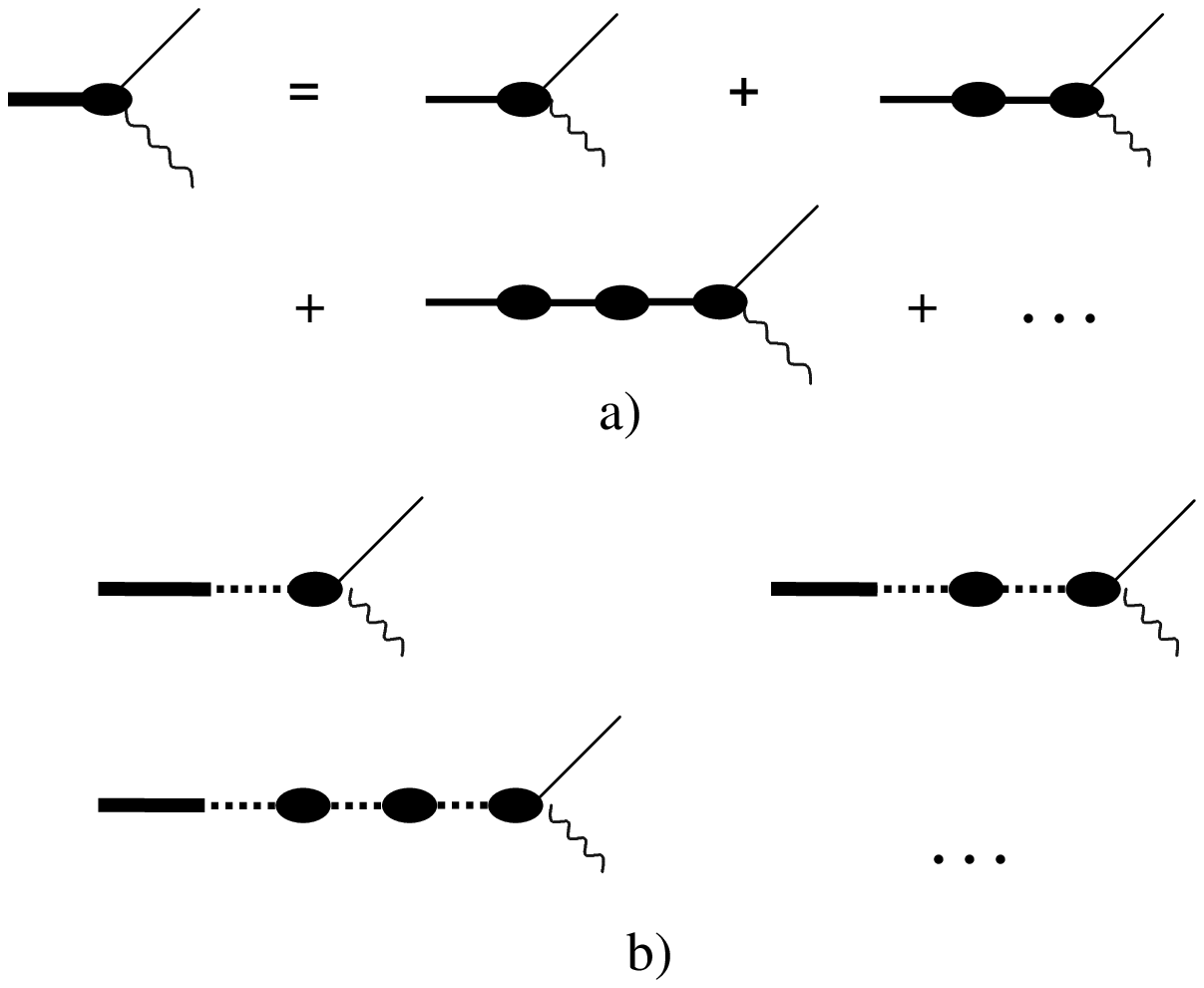}
\begin{minipage}{5.5in}
\caption{Diagrams for $\rho^0\to\pi^0\gamma$ decay. Here ``$\bullet$''
denotes all potential meson loop correction, bold solid lines and dash
lines denote type I complete propagators of $\rho^0$ and $\omega$, which
is defined in fig.~1, ``$\cdots$'' denotes all potential high
order diagrams. In addition, the bold bold solid line in fig.~2-b) is
defined as complete propagators (not type I) of $\rho^0$ in fig.~2-a).
Thus the chain apporximation in fig.~a) corresponds to
renormalization of mass and wave function of external line $\rho^0$, and
the chain apporximation in fig.~b) corresponds to one of internal line
$\omega$.}
\end{minipage}
\end{figure}

There are two kinds of contributions to $\rho^0\rightarrow\pi^0\gamma$
decay. They are shown in the fig.~2-a) and 2-b), and
respectively correspond to nonresonant contribution and
contribution of resonance exchange. Renormalization of mass and wave
function of external line $\rho^0$ is present in both of fig.~2-a) and
2-b), while one of internal line $\omega$ is present in fig.~2-b) only.
For external boson line, Dyson has shwon at adiabatic limit\cite{Dyson51},
\begin{eqnarray}\label{a7}
(q^2-m^2)\Delta_{_F}(q^2)|_{q^2=m^2}\longrightarrow iZ_3^{1\over 2},
\end{eqnarray}
where $\Delta_{_F}(q^2)$ denotes chain approximation to the exact
propagator, in which the mass renormalization has been performed. This
relation hold all of external $\rho^0,\;\pi^0$ and photon fields.
Meanwhile, the renormalization of mass and wave function of internal line
$\omega$ yields its complete propagator as follow
\begin{eqnarray}\label{a8}
\Delta_{(\omega)\mu\nu}^{(c)}(q^2)=\frac{-ig_{\mu\nu}}
  {q^2-m_\omega^2+i\Gamma_\omega(q^2)\sqrt{q^2}}.
\end{eqnarray}
Here the dynamical width $\Gamma_\omega(q^2)$ is generated by pion loops.
It can be also determined by unitarity of $S$-matrix.

According to the above discussion, the LSZ reducation formula~(\ref{a5})
becomes
\begin{eqnarray}\label{a9}
<\pi^0(k)\gamma(q_1),out|\rho^0(q_2),in>=
if_{\rho^0\pi^0\gamma}^{(c)}(q_2^2)|_{q_2^2=m_\rho^2}
\ep^{\mu\nu\alpha\beta}q_{1\mu}\ep_\nu^*q_{2\alpha}e_\beta,
\end{eqnarray}
where the superscript ``(c)'' denotes non-perturbative coupling,
\begin{eqnarray}\label{a10}
f_{\rho^0\pi^0\gamma}^{(c)}(q^2)&=&\bar{f}_{\rho^0\pi^0\gamma}(q^2)
  +\frac{\Pi_{\rho\omega}(q^2)\bar{f}_{\omega\pi^0\gamma}^{(0)}(q^2)}
   {q^2-m_\omega^2+i\sqrt{q^2}\Gamma_\omega(q^2)}.
\end{eqnarray}
In eq.~(\ref{a10}), $\bar{f}_{\rho^0\pi^0\gamma}(q^2)$ and
$\bar{f}_{\omega\pi^0\gamma}(q^2)$ denote renormalized form
factors (momentum-dependent coupling), which include meson loop
correction. Labeling them by a superscript ``(0)'' and taking
$q^2$ to mass shell of $\rho$, eq.~(\ref{a10}) is just the first
equation of eq.~(\ref{3}). The similar discussion holds for the
second equation of eq.~(\ref{3}). So that we finish the prove for
eq.~(\ref{3}) in effective field theory formalism.

\subsection{Mixed propagator approach}

Alternatively, there is another well-known quantum mechanics
method to deal with the $\rho-\omega$ mixing problem:  the
approach of mixed propagator\cite{CB87,renard}. This approach was
developed even before discovery of QCD. The vector meson
propagator is given by (Renard representation)
\begin{eqnarray}\label{aa2}
D_{\mu\nu}(q^2)&=&\int
d^4xe^{-iq.x}\langle 0|T\{V_\mu(x)V_\nu(0)\}|0\rangle \nonumber \\
   &=&D(q^2)g_{\mu\nu}+{1 \over q^2}(D(0)-D(s))q_\mu q_\nu
\end{eqnarray}
where $s\equiv q^2$, and the propagator function $D(s)$ is
written in the following way
\begin{equation}\label{aa3}
D(s)={1\over s-W(s)}.
\end{equation}
For multi-vector-meson channels, $W(s)$ is complex mass-square
matrix with non-zero off-diagonal elements in general.

In order to define physical states measured by experiment, let us
consider the single vector meson resonance channel case first.
The $D_{\mu\nu}$ of eq.~(\ref{aa2}) is now the ordinary propagator
of a vector meson. The reaction amplitude for a process by the
medium of this vector meson resonance reads
\begin{equation}\label{aa11}
{\cal M}\sim J_1^\mu D_{\mu\nu}J_2^\nu=(J_1\cdot J_2){1\over
s-W(s)}
\end{equation}
where $J_1^\mu$ and $J_2^\nu$ represent some currents and $q_\mu
J_{1,\; 2}^\mu=0$. The reaction probability is
\begin{equation}\label{aa12}
\sigma (s)\sim |{\cal M}|^2\propto |{1\over s-W(s)}|^2
\end{equation}
The meson resonance mass measured in the experiment is real and
is determined by the location of the maximum of $\sigma(s)-$peak
in real $s-$axis. Then the resonance mass is determined by the
following equation
\begin{equation}\label{aa13}
{\pa \over \pa s} |{ s-W(s)}|^2=0
\end{equation}
Consequently, the mass-square of the resonance, $M^2$, is
determined by the solution of above equation, i.e.,
\begin{equation}\label{aa14}
s=[ReW-{1\over 4}{\pa\over \pa s}(W^*W+WW^*)](1-{\pa\over \pa
s}ReW)^{-1}\equiv M^2(s).
\end{equation}

Now return to $\rho^0-\omega$ two channel case, $W(s)$ and hence
$M^2(s)$ are complex and real $2\times 2$ matrices respectively.
The mass determination equation for physical resonance states
should read
\begin{equation}\label{aa15}
det[s-M^2(s)]=0.
\end{equation}
The physical states then are the eigenvectors of the real
mass-matrix $M^2$. Following refs\cite{CB87,renard} and using
Breit-Wigner approximation, we have
\begin{eqnarray}
W=\left(
\begin{array}{lcr}
m_{\rho_I}^2-i\sqrt {s} \Gamma_{\rho_I} & \Pi_{\rho\omega}(s) \\
\Pi_{\rho \omega}(s) & m_{\omega_I}^2-i \sqrt{s} \Gamma_{\omega_I}
\end{array}
\right ) \nonumber
\end{eqnarray}
where $m_{\rho_I}$ and $m_{\omega_I}$ are masses of $\rho$ and
$\omega$ in isospin basis. Considering Im$\Pi_{\rho \omega}$ is
small and hence ignorable\cite{OC98}, we get
 \begin{eqnarray}
 ReW=\left(
 \begin{array}{lcr}
m_{\rho_I}^2 & \Pi_{\rho\omega}(s)  \\
\Pi_{\rho \omega}(s) & m_{\omega_I}^2
\end{array} \right ) \nonumber
 \end{eqnarray}
Then, in $\pa ReW/ \pa s$,  only off-diagonal elements of
${\partial \Pi_{\rho \omega}/ \partial s}$ left. Generally, for a
broad class of models $\Pi_{\rho\omega}(s)$ at
$(\rho-,\omega-)$resonance energy region can be determined by
taking VMD-type $\rho-\omega$ mixing Lagrangian  ${\cal
L}_{\rho\omega}=f_{\rho\omega}\rho^{\mu\nu}\omega_{\mu\nu}$ (
$V^{\mu\nu}=\pa^\mu V^\nu-\pa^\nu V^\mu,\ V=\rho,\omega$). It
leads to $\Pi_{\rho\omega}(s)=f_{\rho\omega}s$ which satisfies
$\Pi(s=0)_{\rho\omega}=0$ required by generic consideration in
ref.\cite{O'Connell94}. Thus, we have
\begin{equation} \label{17}
{\partial \over {\partial s}}\Pi_{\rho \omega}=f_{\rho\omega}
    ={\Pi_{\rho \omega} \over s}|_{s\sim m^2_{\rho}}\simeq 6.7\times 10^{-3} \ll 1
   \ \Longrightarrow\ 1-({\partial ReW}/{\partial s})\simeq 1.
\end{equation}
where $|\Pi_{\rho\omega}| \simeq 4000MeV^2 $\cite{O'Connell94} has
been used in the estimation. Furthermore, noting
$\Gamma_{\omega}/m_{\omega}\ll 1$, we have
\begin{eqnarray}
 M^2 = \left(
\begin{array}{cc}
m_{\rho_I}^2-{\partial(s \Gamma_{\rho}^2) \over \partial s} &
     \Pi_{\rho \omega}-{1\over2}(m_{\rho_I}^2+m_{\omega_I}^2)
     {\partial \over \partial s} \Pi_{\rho \omega} \\
\Pi_{\rho \omega}-{1\over2}(m_{\rho_I}^2+m_{\omega_I}^2){\partial
\over \partial s}
    \Pi_{\rho \omega} &
       m_{\omega_I}^2
\end{array}
\right)
\end{eqnarray}
where the off-diagonal elements of $M^2$ matrix represent the
$\rho-\omega$ mixing in the isospin basis. The physical $\rho$
and $\omega$ are eigenstates of $M^2$. In other words, $M^2$
matrix can be diagonalized by the unitary $2\times2$ matrix $C$:
\begin{eqnarray}
C M^2 C^\dag =\left(
\begin{array}{lcr}
m_{\rho}^2  & 0 \\
0 & m_{\omega}^2
\end{array}
\right )\nonumber
\end{eqnarray}
where
\begin{eqnarray}\label{18}
C=\left(
\begin{array}{lcr}
1& -\eta \\
\eta & 1
\end{array}
\right),\hspace{0.5in} \eta = -{\Pi_{\rho \omega}(1-{1\over 2s}
    (m_{\rho}^2+m_{\omega}^2)) \over (m_{\omega}^2-m_{\rho}^2)}.
\end{eqnarray}
Consequently, the solutions physical state condition of
eq.~(\ref{aa15}) are follows
\begin{eqnarray}\label{501}
|\rho_p^0\rangle&=&|\rho_I^0\rangle -\eta |\omega_I\rangle,
\;\;\langle {\rho}^0_p|=|\rho_p^0 \rangle^\dag
 \nonumber \\
|\omega_p\rangle&=&|\omega_I\rangle +\eta |\rho_I^0\rangle,
\;\;\langle \omega_p|=|\omega_p \rangle^\dag.
\end{eqnarray}
Under this transformation, we have
\begin{eqnarray}\label{19}
CWC^{\dag}& = & \left(
\begin{array}{lcr}
 z_{\rho}& T\\
T & z_{\omega}
\end{array}
\right )
\end{eqnarray}
where $z_{\rho}=m_{\rho}^2-im_{\rho}\Gamma_{\rho},z_{\omega}=
m_{\omega}^2-im_{\omega}\Gamma_{\omega}$, and $T=\Pi_{\rho \omega}
-\eta(z_{\omega}-z_{\rho})$ are all defined in the physical state
basis. The propagator function in the physical basis $D^{P}$ reads
\begin{eqnarray}\label{191}
D^P(s) &=&
C(s-W)^{-1}C^{\dag}=(s-CWC^{\dag})^{-1} \nonumber \\
 & = & \left(
\begin{array}{lcr}
(s-z_{\rho})^{-1}& (s-z_{\rho})^{-1} T (s-z_{\omega})^{-1}\\
(s-z_{\omega})^{-1} T(s-z_{\rho})^{-1} & (s-z_{\omega})^{-1}
\end{array}
\right )\nonumber \\
&\equiv&\left(
\begin{array}{lcr}
D^P_{\rho\rho} &D^P_{\rho\rho}T D^P_{\omega\omega}\\
D^P_{\omega\omega} T D^P_{\rho\rho} & D^P_{\omega\omega}
\end{array}
\right ).
\end{eqnarray}
For $V-X$-vertex ($V=\rho,\;\omega$ and $X$ represents other
particles), $f^F_{VX}$
 denotes the corresponding form-factor, $f^P_{VX}$ and $f^0_{VX}$
 denote the coupling constants in the physical basis and in the
 isospin basis respectively.
\begin{figure}[hptb]
\centering
   \includegraphics[width=3.5in]{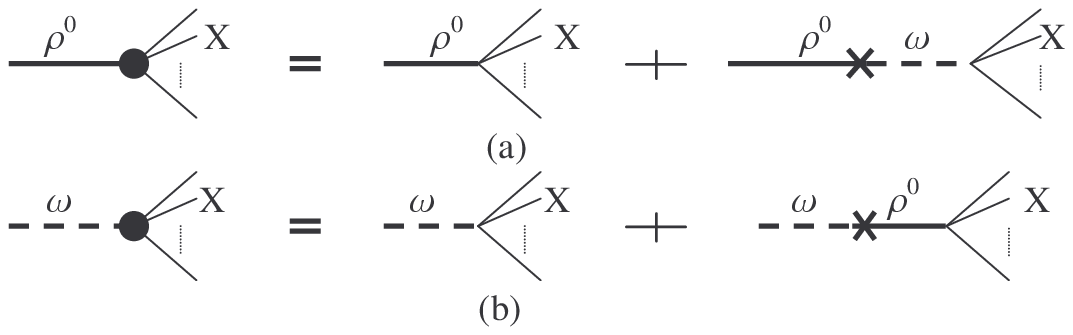}
\begin{minipage}{5.5in}
  \caption{The relation between the form factor of $V-X$ vertex and the corresponding
  coupling constants for the vertices. The black dots in the vertex denotes the form factor.
    The single thick (dash) lines denote $\rho-$propagator $D^P_{\rho\rho}$
   ($\omega-$propagator $D^P_{\omega\omega}$), and the thick-cross-dash
   lines (or dash-cross-thick) line denote the mixed propagator
    $D^P_{\rho\omega}$ (or $D^P_{\omega\rho}$). The thin lines are external lines of
    $X$-particles.}
 \end{minipage}
\end{figure}
Since generally $D^P_{\rho\omega}=D^P_{\omega\rho}\not= 0$, the
form factor is different from the corresponding coupling
constant, i.e., $f^F_{VX}\not=f^P_{VX}$. From Fig.1, we have
\begin{equation}\label{ffactor}
D^P_{VV}f^F_{VX}=D^P_{V\rho}f^P_{\rho X}+D^P_{V\omega}f^P_{\omega
 X},
\end{equation}
with
\begin{eqnarray}\label{602}
f^{P}_{\rho X}&=&f_{\rho X}^{(0)}-\eta
  f_{\omega X}^{(0)}\hspace{1in}
f^{P}_{\omega X}=f_{\omega X}^{(0)}+\eta
  f_{\rho\gamma}^{(0)}.
\end{eqnarray}

In terms of eqs.~(\ref{191}) and (\ref{602}) the time-like EM pion
form-factor is given, in the $\rho-\omega$ interference region, by
\begin{eqnarray}\label{603}
F_\pi (s)&=&1+[f^{P}_{\rho\gamma}D^P_{\rho\rho}f^{F}_{\rho\pi\pi}
+f^P_{\omega\gamma}D^P_{\omega\omega}f^F_{\omega\pi\pi}]\nonumber \\
&=& 1+{f^P_{\rho\gamma} f^P_{\rho\pi\pi}\over s-z_{\rho}}
    +{f^P_{\omega\gamma}f^P_{\omega\pi\pi}\over s-z_{\omega}}
    +{(f^P_{\rho\gamma}f^P_{\omega\pi\pi} + f^P_{\omega \gamma}f^P_{\rho\pi\pi})T
        \over (s-z_{\rho})(s-z_{\omega})}\nonumber \\
  &=&1+(f^P_{\rho\gamma} f^P_{\rho\pi\pi}+{(f^P_{\rho\gamma}f^P_{\omega\pi\pi}
    + f^P_{\omega \gamma}f^P_{\rho\pi\pi})T \over z_{\rho}-z_{\omega}})
   ({1 \over s-z_{\rho}}+\xi e^{i \phi}{1 \over s-z_{\omega}}),
\end{eqnarray}
with
\begin{eqnarray}\label{604}
\xi e^{i\phi}&=&[{1\over 3}\eta-{{{(\eta+{1\over 3})T}\over
{z_\rho -z_\omega}}}][1+{{(\eta+{1\over 3})T}\over {z_\rho
-z_\omega}}]^{-1},
\end{eqnarray}
where $f^{(0)}_{\rho\gamma}=3f^{(0)}_{\omega\gamma}$ and
$f^{(0)}_{\omega\pi\pi}=0$ have been used, and $\phi$ is Orsay
phase. Using $\Pi_{\rho\omega}\simeq -4000MeV^2$\cite{O'Connell94}
in eq.~(\ref{604}), we obtain that $\xi \simeq 0.012$ and $\phi$
is equal to about $100^o -- 101^o$ as $s$ varies from $m_{\rho}^2$
tp $m_{\omega}^2$. These predictions are in good agreement with
experimental data\cite{BEMO98,OC98}, and hence the mixed
propagator approach is legitimate to describe the $\rho-\omega$
mixing effects in the pion EM-form factor.

Now, we study the anomalous-like $\rho^0 \rightarrow \pi^0
\gamma$ and $\omega \rightarrow \pi^0 \gamma$ decays in terms of
the mixed propagator approach. Namely, taking $X=\pi^0\gamma$ in
eq.~(\ref{ffactor}), we have
\begin{eqnarray} \label {701ff}
f^F_{\rho^0\pi^0\gamma} &=&f_{\rho^0\pi^0\gamma}^{(0)}
    -\eta f_{\omega \pi^0\gamma}^{(0)}
    +{T \over m_{\rho}^2-m_{\omega}^2
    +im_{\omega}\Gamma_{\omega}}f_{\omega \pi^0\gamma}^{(0)},\nonumber \\
f^F_{\omega\pi^0\gamma} &=&f_{\omega \pi^0\gamma}^{(0)}
    +\eta f_{\rho^0\pi^0\gamma}^{(0)}
    +{T \over m_{\omega}^2-m_{\rho}^2
    +im_{\rho}\Gamma_{\rho}}f_{\rho \pi^0\gamma}^{(0)},
\end{eqnarray}
where eqs.(\ref{501}) and (\ref{191}) have been used. Considering
$|\eta|\simeq 0.006<<1$ and $T\simeq \Pi_{\rho\omega}$, we
finally obtain eq.(\ref{3}) in mixed propagator formalism.

\section{Numeric result and Conclusion}

The $\omega\rightarrow\pi^+\pi^-$ decay suggests that the on-shell
amplitude $\Pi_{\rho\omega}(m_\rho^2)$ is around
-4000MeV$^2$ which is indeed very small for the most isospin conservation
processes. However, for $\rho^0\rightarrow\pi^0\gamma$ we can see that
$m_\rho^2-m_\omega^2+im_\omega\Gamma_\omega\simeq (-18624+6577 i)$MeV$^2$
is also small due to narrow width of $\omega$. Taking
$\Pi_{\rho\omega}(m_\rho^2)\simeq -4000$MeV$^2$ and combining with
eq.~(\ref{2}), we have
$f_{\rho^0\pi^0\gamma}/f^{(0)}_{\rho^0\pi^0\gamma}\simeq 1.6$ at large
$N_c$ limit and chiral limit. Therefore, the hidden isospin breaking
process $\rho^0\rightarrow\omega\rightarrow\pi^0\gamma$ indeed plays
significant role in $\rho^0\rightarrow\pi^0\gamma$ decay. In addition,
$f_{\omega\pi^0\gamma}/f^{(0)}_{\omega\pi^0\gamma}\simeq 0.99$ is
obtained, so that the contribution from $\rho^0$ exchange in $\omega$
decay can be omitted due to $\Gamma_\rho\gg\Gamma_\omega$ and eq.~(\ref{2}).

For predicting branching ratio for $\rho^0$ decay precisely, the
precise on-shell $\rho^0-\omega$ mixing amplitude is needed. The
investigation of $\rho^0-\omega$ mixing has been an active
subject\cite{CB87,BEMO98,OC98,WY00}. In ref.\cite{OC98} the
on-shell mixing amplitude has been determined as
$\Pi_{\rho\omega}(m_\omega^2)=-[(3500\pm 300)+(300\pm
300)i]$MeV$^2$ in the pion form factor in the timelike region, but
in this fit effect of ``direct'' $\omega\pi\pi$ coupling is
omitted. We quote this result as $R_1$. Meanwhile, the
ref.\cite{WY00} provided a complete theoretical study on this
mixing, which included the effect of ``direct'' $\omega\pi\pi$
coupling, and is up to the next to leading order of $N_c^{-1}$
expansion. The result is $\Pi_{\rho\omega}(m_\omega^2)=-[(3956\pm
280)+(1697\pm 130)i]$MeV$^2$, and we quote this result as $R_2$.
Then using $B(\rho^\pm\rightarrow\pi^\pm\gamma)=(4.5\pm
0.5)\times 10^{-4}$ and
$B(\omega\rightarrow\pi^0\gamma)=(8.5\pm 0.5)\times 10^{-2}$
and assuming
$f^{(0)}_{\rho^0\pi^0\gamma}=f_{\rho^\pm\pi^\pm\gamma}$, we obtain~
\footnote{This result includes meson loop correction, since here we using
$B(\rho^\pm\rightarrow\pi^\pm\gamma)$ and
$B(\omega\rightarrow\pi^0\gamma)$ to fit $f_{\rho^0\pi^0\gamma}$ and
$f_{\omega\pi^0\gamma}^{(0)}$ instead of using eq.~(\ref{2}).}
\begin{eqnarray}\label{4}
B(\rho^0\rightarrow\pi^0\gamma)=\left\{ {{\displaystyle (11.05\pm
1.84)\times 10^{-4}\hspace{1in}{\rm for}\;\;R_1}\atop
{\displaystyle (12.25\pm 1.52)\times 10^{-4}\hspace{1in}{\rm for}\;\;R_2}}
\right.
\end{eqnarray}
Here we have used $\Pi_{\rho\omega}(m_\rho^2)\simeq
m_\rho^2\Pi_{\rho\omega}(m_\omega^2)/m_\omega^2$. The above results
strongly support the fit of the solution A in table 1 instead of the
solution B. In fact, from viewpoint of experiment, the solution A is also
better than the solution B (because of smaller $\chi^2$ and reasonable
phase difference). Therefore, we conclude that the branching ratio for
$\rho^0\rightarrow\pi^0\gamma$ is
\begin{eqnarray}\label{6}
B(\rho^0\rightarrow\pi^0\gamma)=(11.67\pm 2.00)\times 10^{-4}.
\end{eqnarray}
Here our phenomenological estimates (eq.~(\ref{4})) are kept as
references instead of averaging them and experimental fit.

Finally, it is interesting to estimate the contribution of
$\omega$ exchange in $\rho^0\rightarrow\eta\gamma$ decay. This
process itself is isospin breaking due to electromagnetic
interaction. The exact isospin symmetry implies
$f^{(0)}_{\rho^0\eta\gamma}/f^{(0)}_{\omega\eta\gamma}=3$. Hence
similar argument gives
$f_{\rho^0\eta\gamma}/f^{(0)}_{\rho^0\eta\gamma}\simeq 1.06$ and
$B(\rho^0\rightarrow\eta\gamma)/B(\omega\rightarrow\eta\gamma)\simeq
0.5$. This result agrees with current experimental fits,
$B(\rho^0\rightarrow\eta\gamma)=(2.4{+0.8\atop -0.9})\times
10^{-4}$ and $B(\omega\rightarrow\eta\gamma)=(6.5\pm 1.0)\times
10^{-4}$\cite{PDG2000}.

To conclude, we show that a hidden effect of isospin symmetry
breaking plays essential role in $\rho^0\rightarrow\pi^0\gamma$
decay. The transition amplitude is derived by effective field
theory approach and mixed propagator approach respectively. The
result yielded by two independent method is matched each other.
Our result is also supported by some experimental evidences. It
indicates that the current datum for this process in PDG should
be corrected.


\begin{thebibliography}{99}
\bibitem{Miller90}G.A. Miller, B.M.K. Nefkens and I. $\check{\rm S}$laus,
Phys. Rep. {\bf 194}, 1 (1990).
\bibitem{PDG2000}Particle Data Group, D.E. Groom {\sl et al.}, Eur. Phys.
J. {\bf C 15}, 409 (2000) .
\bibitem{ccho}T. Jensen {\sl et al.}, Phys. Rev. {\bf D 27}, 26 (1983);
J. Huston {\sl et al.}, Phys. Rev. {\bf D 33}, 3199 (1986);L. Capraro {\sl
et al.}, Nucl. Phys. {\bf B 288}, 659 (1987).
\bibitem{Dru84}V.P. Druzhinin {\sl et al.}, Phys. Lett. {\bf B 144}, 136
(1984).
\bibitem{Dolinsky89}S.I. Dolinsky {\sl et al.}, Z. Phys. {\bf C 42}, 511
(1989).
\bibitem{Dolinsky91}S.I. Dolinsky {\sl et al.}, Phys. Rep. {\bf 202}, 99
(1991).
\bibitem{Benayoun96}M. Benayoun S.I. Eidelman and V.N. Ivanchenko, Z.
Phys. {\bf C 72}, 221 (1996).
\bibitem{Benayoun93}M. Benayoun {\sl et al.}, Z Phys. {\bf C 58}, 31
(1993).
\bibitem{Benayoun95}M. Benayoun {\sl et al.}, Z Phys. {\bf C 65}, 399
(1995).
\bibitem{LSZ55}H. Lehmann, K. Symanzik and W. Zimmermann, Nouvo
Cimento, 1 (1955) 205.
\bibitem{WY00}X.J. Wang and M.L. Yan, Phys. Rev. {\bf D 62}, 094013
(2000);{\sl Chrial Expansion Theory at Vector Meson Scale}, X.J. Wang
and M.L. Yan, hep-ph/0010215.
\bibitem{Dyson51}F.J. Dyson, Phys. Rev. {\bf 83} (1951) 608.
\bibitem{CB87}Coon,S.A., Barrett,R.C.: Phys. Rev. {\bf C36}, 2189 (1987);
T. Goldman, J.A. Henderson and A.W. Thomas, Few Body Systems {\bf
12}, 123 (1992);A. Bernicha, G. Lopez Castro and J. Pestieau,
Phys. Rev. {\bf D 50}, 4454 (1994);K. Maltman, H.B. O'Connell and
A.G. Williams, Phys. lett. {\bf B 376}, 19 (1996);R. Urech, Phys.
Lett. {\bf B 355}, 308 (1995);H.B. O'Connell, A.W. Thomas and A.G.
Williams, Nucl. Phys. {\bf A 623}, 559 (1997);
\bibitem{renard} F.M.Renard, Springer Tracts in Modern Phys., {\bf
63}, 98-120, Springer Verlag (1972).
\bibitem{BEMO98}M. Benayoun, S. Eidelman, K. Maltman, H.B. O'Connell,
B. Shwartz and A.G. Williams, Eur. Phys. J. {\bf C 2} 269 (1998).
\bibitem{OC98}S. Gardner and H.B. O'Connell, Phys. Rev. {\bf D 57}, 2716
(1998).
\bibitem{O'Connell94}H.B. O'Connell, B.C. Pearce, A.W. Thomas and
A.G. Williams, Phys. Lett. {\bf B 336}, 1 (1994).

\end{thebibliography}
\end{document}